# Observational methods for solar origin diagnostics of energetic protons

Rositsa Miteva

Space Research and Technology Institute – Bulgarian Academy of Sciences

rmiteva@space.bas.bg

**Abstract**

The aim of the present report is to outline the observational methods used to determine the solar origin – in terms of flares and coronal mass ejections (CMEs) – of the in situ observed solar energetic protons. Several widely used guidelines are given and different sources of uncertainties are summarized and discussed. In the present study, a new quality factor is proposed as a certainty check on the so-identified flare–CME pairs. In addition, the correlations between the proton peak intensity and the properties of their solar origin are evaluated as a function of the quality factor.

**Introduction**

The Sun is a constant source of electromagnetic radiation covering almost the entire spectrum. The highest energy wavelengths observed (hard X-rays and gamma-rays) are produced during a specific type of eruptive events named solar flares (Fletcher et al. 2011). Although the flares are equated with the 'flash of light' observed in a specific wavelength range by the observing instrument, they are also a process that involves mass motion (jets), magnetic field restructuring and acceleration of particles. The underlying physical mechanism that drives the solar flare is called magnetic reconnection. Despite that the details of the process need to be better understood, the term is well adopted to qualitatively explain the various eruptive phenomena in the solar corona. A different manifestation of solar activity is the coronal mass ejection (CME). This is an enormous bubble of plasma and embedded magnetic field (Chen, 2011) that escapes the solar pull and continues to expand during their propagation into the interplanetary (IP) space. Ahead of the CME, the plasma and magnetic field are compressed and a turbulent region is formed with suitable conditions for local acceleration of particles (both in terms of reconnection with coronal/IP field lines or due to shock acceleration). Thus both flares and CMEs are regarded as the probable drivers (the solar origin) of energetic particles.

Solar energetic particles (SEPs) is the term used nowadays to describe the population of protons, ions and electrons with energies from keV to MeV that is observed in situ (Schwenn, 2006, Desai and Giacalone, 2016). Presently, the majority of the satellites equipped with particle detectors are situated at L1 (since 1996) together the twin-STEREO mission (since late 2006) orbiting the Sun from opposite directions at a distance of 1AU and in the ecliptic plane.

The interest to study the SEP events is justified due to the risk they pose to technological devices and satellites as a whole. Energetic protons, in particular, can give rise to dangerously elevated radiation doses of astronauts during their time in space. Together with solar flares and CMEs, the SEP events constitute the important components of space weather and are subject to active research (Pulkkinen, 2007).

Great efforts are directed towards the successful forecasting of energetic particles (mostly protons). Numerous schemes are proposed to forecast the proton event on first place, and then their arrival time and the maximum proton intensity (e.g, Nunez 2011). Irrelevant on the methodology used, the forecasts need to be validated against real observations, thus comprehensive lists of observed particle events are needed. Since the observing instruments are subject to data gaps, several independent satellites are needed to compile a consistent particle list. A number of them are already available, see a discussion in Miteva et al. (2017).

After the in situ identification of the particle event (as an enhancement of proton or electron intensity above the background level), the attention is turned toward the Sun and the eruptive phenomena which occurred there. Very often, before the particle enhancement, a pair of a flare and a CME can be identified. The profiles of the particle events are indicative of the helio-location of the active region (AR) where flares/CMEs originate, ranging from western or eastern (Lario and Simnett, 2004). During high solar activity periods (solar maximum or/and high productivity phase of an AR) there are multiple flare-CME pairs and the allocation of the particle event to a specific pair becomes subjective. Moreover, the flare-accelerated particles need an escape route through the solar corona, otherwise confined configurations are formed. The broad CME shock fronts are often regarded to be free from such limitations, provided the shock acceleration is effective over the entire shock front which is an overestimation. In either case, a magnetic field line connection between the acceleration site and the detector must be established and kept in time for the gyrating particles to be finally observed.

The long-standing issue about the identification of the origin of SEP events is not yet been solved. There are numerous reasons, among them: a wide range in time delays between the remote observation of flares/CMEs and the in situ detection of particles; the observed properties of the flares/CMEs are subject to occultation/projection effects, respectively; a single (taken as a representative) value of the flare/CME phenomena is adopted and correlated with the particle peak intensity using the simplest correlation analysis possible usually without filtering out well-known inter-correlations; time dependent effect of acceleration efficiency and injection of particles; scattering of the particles in the IP space.

Data collected from the new generation spacecraft also showed that the particles form a continuous distribution in various properties (e.g, Cane et al. 2010) and a simple bi-modal division (Reames (1999) with subsequent revisions) is the exception, not the rule. In summary, there seems to be a gradual shift in view that it is not 'either' but actually 'both' flares and CMEs contribute to the particle enhancements observed in situ, although some impulsive opinions continue to resist. The degree of the individual contribution of flares and CMEs to the SEP events, both as a function of time and location, is still to be defined in a quantitative way.

**Solar origin association procedure**

For the present report, a list of 20 MeV proton events from SOHO/ERNE instrument (Torsti et al. 1995) is used. A preliminary analysis, based on 5-minute smoothed data is completed. For the analysis presented here, only the events during solar cycle (SC) 23 will be used: namely in the period 1996–2008. The complete event list is being developed under an ongoing project (http://newserver.stil.bas.bg/SEPorigin). A proton event is identified when the proton intensity rises above a specific level, selected here at three standard deviations on the background level (the latter is calculated as the average intensity over a quiet period selected by the observed). Finally, the peak proton intensity is the amplitude between so-identified pre-event background and the peak value on the profile (proton enhancements due to IP shocks are not considered). In the considered period of interest there are 373 proton events.

Based on timing arguments, the SEP-related flare and CME is also identified. Namely, the strongest flare and the fastest and widest CME are preliminary selected as the solar origin. Depending on the rise time (onset to peak), a western flare/CME is preferred for rapidly rising proton profiles or alternatively, an eastern pair – in case of a slowly rising profile. For the flare and CME to constitute a pair, they need to commence in a time period of about one hour and the helio-location of the flare AR and the direction of propagation of the CME need to be in the same solar disk quadrant. Some well-known catalogs used for flare (ftp://ftp.ngdc.noaa.gov/STP/space-weather/solar-data/solar-features/solar-flares/x-rays/goes/ and www.solarmonitor.org) and CME

(https://cdaw.gsfc.nasa.gov/CME_list/) properties were used.

Here, a quality estimation for the degree of certainty on the association of the SEP-producing flare–CME pair is proposed. The certainty domain is roughly divided into three parts:

- a high degree of certainty, when prior to the proton enhancement a single flare–CME pair is present or, alternatively, when other pairs are present they are about 1–2 hours apart in onset to the primary pair selected;
- an average degree of certainty, when multiple flare–CME pairs are observed, but the selected one is the strongest – in terms of flare class and CME speed, and also in the case when either flare or CME fulfills the above condition but there is a data gap for the other eruptive event;
- low degree of certainty, when multiple flare–CME pairs with comparable strength are noticed in 1–2 hour period, however a weaker pair but closer in time to the proton onset is selected or when a much stronger pair closely followed by a weaker pair is preferred: all these cases are complex events with a rather high degree of subjectivity.

For each case in the proton list when a solar origin association is proposed, the above set of criteria is followed. Note that the criteria is imposed upon the flare–CME pair, and not on either of them separately. The effect of on-sky projection of the CME speed and partially occulted flare emission will not be considered in this study.

**Results**

Out of the entire 373 proton events, no flare could be associated in 109 of the cases (about 29% of the sample), whereas no CME could be related to 68 proton events (18%). From the above, neither flare nor CME could be identified in 42 cases (about 11% of the entire sample) and these events are dropped from further statistical analysis. A distribution with respect to the time is shown in Fig. 1 as stacked histogram, using bins with a width of 6 months. The number of events in each bin is the sum of all color sections, whereas the different colors denote the certainty level. For the entire period, as shown in the plot, with black is denoted the low certainty events (61 cases; 17% of the sample), dark-gray – average (124; 33%), light-gray – high degree of certainty in solar origin identification (146; 39%). The remaining 11% comprise the uncertain case from above.

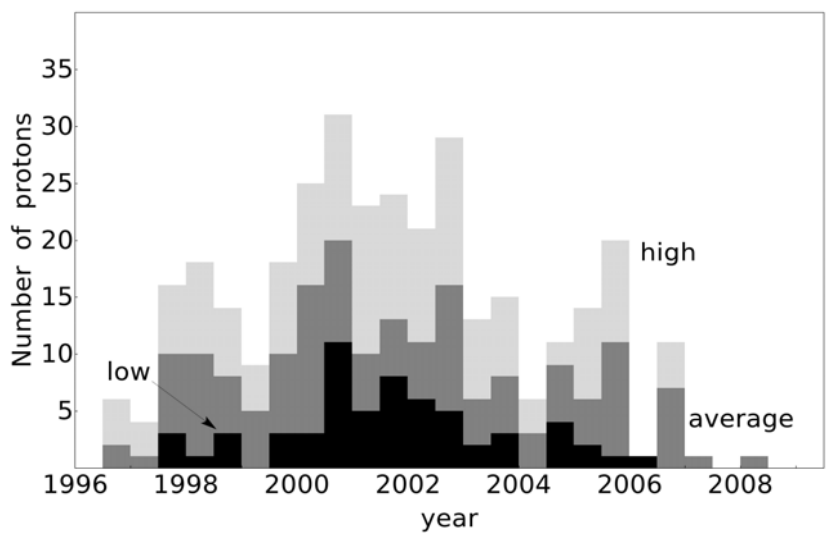

Figure 1. Stacked histogram of the number of proton events over the years shown in 6-month binning. The colors denote the certainty level: light-gray – high, dark-gray – average and black – low certainty. The length of the color bar denotes the number of proton events with identified solar origin

of the specific certainty and the sum of all three sections gives the total number of proton events with solar origin.

It is customary to calculate the Pearson correlation coefficients between the peak proton intensity in units of protons/(cm^2 s sr MeV), with the flare class or/and CME speed. These values were used to argue in favor of one accelerator over the other, often without the effort to evaluate uncertainties (error margin) on the correlation coefficients. For the flares, the peak value in GOES soft X-ray flux (as measured in 1-8 Å channel) is used, termed flare class (i.e., X1 is the flux with a value of $10^{-4}$ W/m^2, and the remaining letters (M, C, B) denote 10 times less flux). For the CME, the reported linear speed is used (projected on the plane of sky). The scatter plots are given in Fig. 2, where the different symbols denote the certainty level (filled black circles are for the high certainty, gray – for average, and empty circles – for low certainty). A large scatter is usually evident in either case. One could argue, that in the case of lowest certainty, the large proton intensity events are systematically related to lower class flares (left plot) and slower CMEs (right plot), thus randomizing the correlations.

The values of all correlation coefficients (in log–log form) between the peak proton intensity with flare class and CME speed are summarized in Table 1, for different levels of certainty. As also seen in Figure 2, the lower level of certainty introduces an additional scatter leading to lower value for the correlations. The largest correlation coefficient in either case (flares and CMEs) is obtained for the sample containing only events with high certainty of solar origin association.

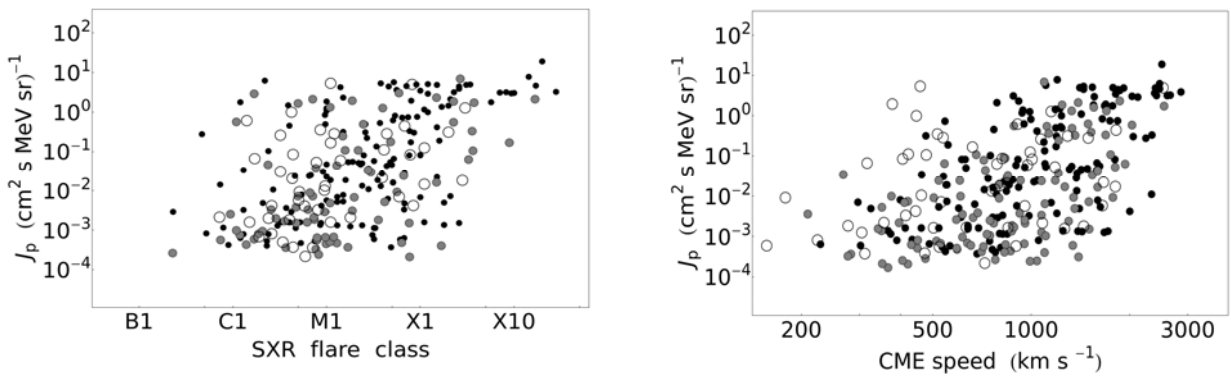

Figure 2. Scatter plots in the log-log format between the peak proton intensity and the flare class (left) or CME speed (right). The filled clack circle is for events with high level of certainty, gray circle is for average, whereas empty circle – for low level of certainty in solar origin identification.

Compared to the entire sample (irrelevant on the level of certainty, see top row in Table 1), a statistically significant result is obtained for the sub-sample with the highest certainty (second row in the Table), but only for the case when correlating with CMEs (0.62±0.05 compared to 0.52±0.04). In contrast, when the correlation is done for the flare class the result is the same (0.53±0.05 compared to 0.51±0.05), respectively. The uncertainties are calculated as described in Miteva et al. (2013). The value of the correlation coefficient for flares/CMEs decreases when the sub-sample of average certainty is included, namely to 0.52/0.58, respectively, with some overlap due to the uncertainty range. When considering only the sub-sample of low certainty (last row in the Table), the correlation is the lowest, 0.40±0.11 (with flares) and 0.32±0.12 (with CMEs), which is an evidence for the scattering. The influence of such spread is the largest when performing correlations with the CME speed, compared to the flare class.

Table 1. Pearson correlation coefficients (log–log) between the peak proton intensity and the flare class (column 2) and CME linear speed (column 3) for samples with different level of certainty. In brackets are given the exact number of events is the specific sub-sample used in the correlation analysis.

| Level of certainty | Correlations with flare class | Correlations with CME speed |
|---|---|---|
| Any (all events) | 0.51±0.05 (264) | 0.52±0.04 (305) |
| High (only) | 0.53±0.05 (145) | 0.62±0.05 (145) |
| High + average | 0.52±0.05 (218) | 0.58±0.04 (251) |
| Average (only) | 0.47±0.11 (73) | 0.48±0.08 (106) |
| Low (only) | 0.40±0.11 (45) | 0.32±0.12 (53) |

**Discussion**

For the first time, a quality factor is set on the certainty of the proposed flare–CME pair association to each proton event. As expected, the lower certainty adds to the scatter in the correlations and reduces their correlation coefficient value. This could be also explained with the fact that in the sample of highest certainty, the selected flares/CMEs occur closest in time before the proton enhancements, compared to the other certainty sub-samples. In such a case, flares and CMEs were not modified substantially in terms of their acceleration potential. Finally, the adopted values for the flare class and CME speed used in the correlation can be regarded as representative ones for the eruptive process under way.

The inclusion of a quality factor on the proposed flare-CME pair is one of the possible methods to evaluate the level of certainty of the SEP origin association. However, a possible step for future improvement of the proposed technique is to assign a quality factor separately for each type of eruptive phenomenon – individually for flares and for CMEs. Moreover, flares and CMEs are subject to different kind of effects, e.g., partially occulted emission for the flares/back-sided locations vs. projection effects for the CMEs, that deserves to be adequately addressed.

In summary, with the proposed quality factor it was shown that subjectivity and erroneously associated solar origin to proton events can account for some reduction in the statistical correlation coefficients, however, the difference is not statistically significant when comparing the correlations between protons with flares vs. protons with CMEs. Other reasons need to be explored, both observationally and theoretically, in order to isolate, if possible, the influence of flares from CMEs to the resultant observed in situ particle flux.

**Acknowledgements**

This study is supported by the project 'The origin of solar energetic particles: solar flares vs. coronal mass ejections' which is co-funded by the National Science Fund of Bulgaria with Contract No. ДНТС/Russia 01/6 (23-Jun-2017) and by the Russian Foundation for Basic Research with Project No. 17-52-18050.